\documentclass[
    aip,
    rsi,
    amsmath,
    amssymb,
    reprint
]{revtex4-1}

\usepackage[utf8]{inputenc}
\usepackage[T1]{fontenc}
\usepackage{graphicx}
\usepackage{hyperref}
\usepackage{etoolbox}  
\usepackage{siunitx}

\makeatletter
\def\@email#1#2{%
 \endgroup
 \patchcmd{\titleblock@produce}
  {\frontmatter@RRAPformat}
  {\frontmatter@RRAPformat{\produce@RRAP{*#1\href{mailto:#2}{#2}}}\frontmatter@RRAPformat}
  {}{}
}%
\makeatother

\DeclareSIUnit\bit{bit}
\DeclareSIUnit\byte{byte}

\begin{document}

\title{Software-defined lock-in demodulator for low-frequency resistance noise measurements}
\author{Tim Thyzel}
\email{thyzel@physik.uni-frankfurt.de}
\affiliation{Physikalisches Institut, Goethe-Universität Frankfurt am Main, Germany}

\date{\today}

\begin{abstract}
The resolution of low-frequency resistance noise measurements can be increased by amplitude modulation, shifting the spectrum of the resistance fluctuations away from the \(1/f\) noise contributed by measurement instruments.
However, commercial lock-in amplifiers used for de-modulating the fluctuations exhibit a problematic \(1/f\) noise contribution, which imposes a hard lower limit on the relative resistance noise that can be detected. 
We replace the lock-in amplifier hardware by equivalent digital signal processing performed using open-source software and off-the-shelf data acquisition systems.
Compared to previous implementations of the lock-in principle, our solution offers real-time preview capabilities and is resource-efficient for long acquisition times at high sampling rates. 
Importantly, compared to high-end commercial lock-in instruments, our system offers superior low-frequency noise performance with a reduction of the voltage power spectral density by about two orders of magnitude.
\end{abstract}

\maketitle

\section{Introduction}

Low-frequency resistance noise measurements detect small fluctuations of the electrical resistance
of a device under test (DUT) in the frequency range of millihertz to kilohertz.
They can offer valuable insights into impurity states in semiconductor devices \cite{Fleetwood2015}, as well as structural \cite{Mueller2015}, magnetic \cite{Weissman1993} and charge carrier dynamics \cite{Mueller2011,Mueller2018} in fundamental condensed matter physics.
For this type of measurement, the resistive DUT is usually excited with a constant current and fluctuations in the voltage drop are detected.
As the relative standard deviation of these fluctuations is often only of order \(\sigma_R / R = O(\SI{e-5})\), the instruments used to measure them, in particular amplifiers, must add as little noise to the experiment as possible.
In particular, \(1/f\) noise caused by the resistive elements (e.g.\ in semiconductor devices) of measurement circuits and instruments is problematic, because it may mask the \(1/f\) noise of the DUT, whose detection is the objective of the measurement.

One way to eliminate instrument \(1/f\) noise is to modulate the DUT resistance with a sinusoidal current, to amplify the signal, and then to de-modulate it using phase-sensitive (``lock-in'') detection.
This AC method \cite{Scofield1987} shifts the power density spectrum of the DUT resistance noise away from the amplifier's \(1/f\) noise contribution.
However, the lock-in amplifier hardware commonly used for demodulation exhibits additional \(1/f\) noise, which furthermore increases with input signal amplitude.
The relative voltage noise power spectral density
\begin{align*}
\frac{S_V(f)}{\langle V \rangle^2} = \frac{2}{T} \Bigg\lvert \int\limits_0^T \frac{\delta V(t)}{\langle V \rangle}\  e^{-2 \pi i f t} \, dt \Bigg\rvert^2
\end{align*}
of this \(1/f\) background has a value of \(S_V(f = \SI{1}{\hertz}) / \langle V \rangle^2 \approx \SI{2e-11}{\per\hertz}\) for the Stanford Research Systems (SRS) SR830 instrument commonly found in physics laboratories.
The quadratic dependence of this background on the mean voltage \(\langle V \rangle\) matches that of the DUT noise voltage drop, for which \(S_V \propto S_R \times \langle I \rangle^2 = (S_R/\langle R \rangle^2) \times \langle V \rangle^2\) according to Ohm's law, making them difficult to distinguish.
Therefore, the measurement resolution is limited if a bridge configuration \cite{Scandurra2022} to suppress \(\langle V \rangle\) is not available.

\begin{figure}
    \centering
    \includegraphics[scale=1.0]{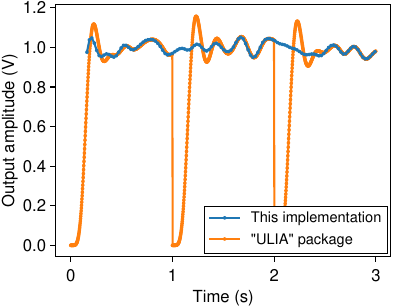}
    \caption{Comparison of the output signal of our software package to the \texttt{ulia} package described in Ref.\ \cite{Uhl2021}.
        A synthetic input signal of a \(\SI{50}{\hertz}\) sine wave superimposed with white noise, divided into consecutive blocks of \(\SI{1}{\second}\) duration, has been demodulated with both software packages.
        The \texttt{ulia} package does not retain filter state across block boundaries, introducing discontinuities undesirable for noise measurements.
        Our implementation yields a smooth, continuously filtered output, which is futhermore downsampled to decrease the data rate.
    }
    \label{fig:ulia-comparison}
\end{figure}

To circumvent this problem, we replace the lock-in amplifier hardware with software digital signal processing.
Our implementation, specifically designed for low-frequency noise measurements, offers advantages in ease of use and resource efficiency over previous reports\cite{Uhl2021,ODonoghue2015,Andersson2007}.
First and foremost, noise measurements usually necessitate the acquisition of signals over durations up to one hour.
If demodulation takes place only after the full input signal has been acquired as in Ref.\ \cite{Uhl2021}, the output cannot be evaluated before the experiment ends. 
Therefore, we implement a streaming architecture where lock-in demodulation takes place as the input data is being acquired, allowing a quasi-real-time preview of the output signal.
This feature has not been previously reported, see Fig.\ \ref{fig:ulia-comparison}, and it makes our solution a drop-in replacement for conventional lock-in amplifier hardware.

Second, the lock-in carrier frequency, and therefore the alias-free sampling rate, is often far higher -- \(\SI{100}{\hertz}\) to \(\SI{100}{\kilo\hertz}\) -- than the frequency scales of interest for noise measurements -- e.g.\ \(\SI{0.1}{\hertz}\) to \(\SI{10}{\hertz}\).
Storing the demodulated output signal at this sampling rate in memory and persistent storage as in Ref.\ \cite{Uhl2021} is inefficient, producing data at a rate of \(O(\SI{1}{\giga\byte})\) per hour when sampling at \(O(\SI{100}{\kilo\hertz})\).
Therefore, the demodulator output should be downsampled/decimated as it is being acquired, leading to a lower output data rate and a smaller memory footprint.
This is advantageous for integrating the lock-in amplifier into a larger signal-processing system, which may perform computationally expensive operations such as Fourier transforms on the demodulated output or archive it in long-term storage.
A previous implementation achieves downsampling by conducting the signal processing on a microcontroller-based hardware system \cite{ODonoghue2015}, which is highly cost-effective, but requires the design and manufacturing of custom electronics.
Our solution makes use of commercial, off-the-shelf data acquisition hardware and implements alias-free, streaming downsampling in software.
Thanks to the careful use of buffers, the software does not require real-time scheduling, and therefore runs on general-purpose computer operating systems.
Furthermore, a modular architecture allows the use of acquisition hardware from different vendors, an advantage over custom electronics in varying laboratory environments.
In the following, the design of this open-source software package, as well as the accompanying hardware setup is described, and a significant improvement in low-frequency noise performance is demonstrated over lock-in amplifier hardware.

\section{Implementation}

\begin{figure}
\raggedright
\hspace{4mm}\includegraphics[scale=0.9]{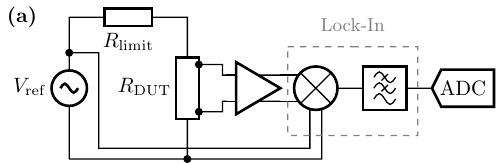}\\
\vspace{3mm}
\hspace{4mm}\includegraphics[scale=0.9]{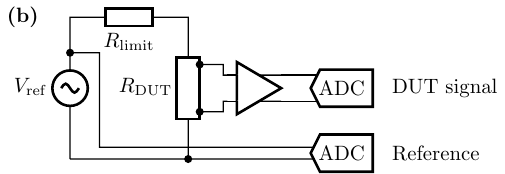}
\caption{\textbf{(a)}: Circuit for AC noise measurements using Lock-in amplifier hardware.
    \textbf{(b)}: Circuit for the software-based technique.
}
\label{fig:circuits}
\end{figure}

Fig.\ \ref{fig:circuits}a shows the circuit used for noise measurements with conventional lock-in amplifier hardware, in which a function generator (e.g.\ SRS DS360) with low harmonic distortion excites the DUT resistance with a current set by \(R_\mathrm{limit}\).
The DUT voltage drop is amplified and passed into the lock-in instrument, which includes a mixer and a low-pass filter, and which outputs the demodulated noise signal to be recorded by an analogue-to-digital converter (ADC).
The noise power spectral density \(S_V\) is then calculated from the recorded time series.
Note that this setup is only sufficient if the DUT resistance noise is not bias-dependent, which needs to be verified in all practical applications using suitable DC methods \cite{Scandurra2022,Thyzel2024Methods}.
For the software-based technique, see Fig.\ \ref{fig:circuits}b, the DUT voltage is buffered by a low-noise, multi-stage amplifier (SRS SR560), whose output is directly recorded by an ADC.
A second, simultaneously sampling ADC records the function generator output \(V_\mathrm{ref}\), which will serve as a reference for the demodulation conducted in software.
The differential, isolated ADCs of the Delta-Sigma-type \cite{Kester2005} are provided by a data acquisition system (DAQ, NI 9239), which features a high resolution (\SI{24}{\bit}), sufficient bandwidth (\SI{25}{\kilo\hertz}) and a low noise floor (\(\sqrt{S_V(\SI{70}{\hertz})} \approx \SI{400}{\nano\volt\per\sqrt\hertz}\)) with an input range of \(\SI{\pm 10}{\volt}\).

\begin{figure*}
    \centering
    \includegraphics[width=\textwidth]{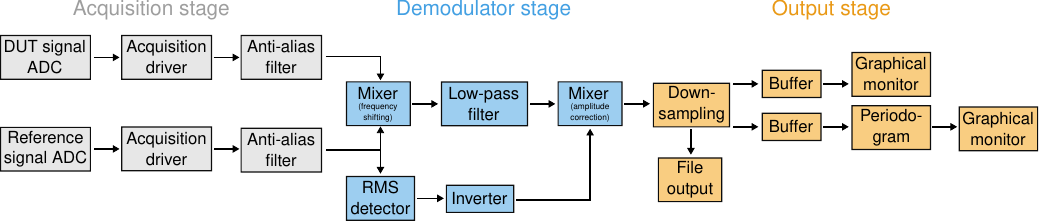}
    \caption{Data flow diagram of the software package.
        Each rectangle (except for the ADCs on the far left) represents a signal processing step implemented by a Python object conforming to the Observable and/or Observer protocol.
        Time series blocks are emitted along each arrow.
    }
    \label{fig:software-diagram}
\end{figure*}

The software signal processing system, which we make available under an open-source license at Ref.\ \cite{Thyzel2024Software}, is visualized in Fig.\ \ref{fig:software-diagram}.
Its acquisition stage employs driver modules, in this case the NI DAQmx framework via the \texttt{nidaqmx} Python package, to set up and read data from the ADCs.
Note that data acquisition hardware from other vendors can be easily integrated by implementing a new acquisition driver module.
Instead of processing samples of the input voltage one-by-one, double buffering allows an uninterrupted, block-wise readout.
This is advantageous because, although the execution speed of Python is limited by the interpreter, a large ecosystem of high-quality scientific libraries (NumPy \cite{NumPy2020}, SciPy \cite{SciPy2020}) is available that deal efficiently with arrayed data, such as time series blocks.
Furthermore, this buffered design is necessary on computer operating systems that do not support real-time scheduling, because it cannot be guaranteed that a single sample will be fully processed before the next one is acquired, as is the case in microcontroller-based implementations such as Ref.\ \cite{ODonoghue2015}.
Block-wise signal processing is implemented using the Observer software architecture pattern \cite{DesignPatterns}, also known as Publish-Subscribe.
Once the acquisition driver, an Observable object, has filled one of its two buffers, it emits the buffer contents as a time series block, which is received by one or more signal processing Observers.
Each Observer object applies a transformation (e.g.\ filtering, mixing, downsampling, ...) to the time series data in the block and emits the result to other Observers further along the chain.
In this manner, a quasi-real-time signal processing pipeline is built by composing Observable and Observer objects.

At the end of the acquisition stage, the first signal processing step applies a low-pass filter on both channels independently to avoid aliasing in the mixing steps to follow.
Butterworth-type filters of sixth order are used, which offer an ideally flat passband important for broad-band noise measurements, and a good roll-off rate of \(\SI{-36}{\deci\bel}\) per octave \cite{OppenheimDsp}.
Their infinite impulse response allows an implementation by a difference equation, which offers higher computational efficiency than filters based on convolution, and which is available through a combination of the functions \texttt{scipy.signal.butter} and \texttt{scipy.signal.sosfilt}.
Filter state is retained across time series blocks using the \texttt{zi} argument, which is vital to avoid introducing discontinuities across block boundaries, as illustrated in Fig.\ \ref{fig:ulia-comparison}.

In the demodulator stage, the pre-filtered DUT and reference signals are mixed, i.e.\ multiplied element-wise.
This operation performs amplitude de-modulation of the DUT signal, shifting the DUT spectrum back to zero frequency.
However, it also introduces a spectrum copy at twice the reference frequency, which needs to be removed by a low-pass filter.
Furthermore, mixing modifies the amplitude of the DUT signal, which is an undesirable side-effect.
To revert this amplitude scaling, an RMS detector comprised of a squaring mixer, a low-pass filter and a square-root operation outputs a DC time series of the root-mean-square amplitude of the reference signal.
The output of the demodulation low-pass filter is then multiplied with the inverse of the reference amplitude to return to the correct DUT signal amplitude.

The time series blocks produced by the demodulator stage are passed to the output stage, where they are down-sampled, or decimated, to reduce the data rate.
Care is taken to maintain correct sample spacing across block boundaries.
The new sampling rate is twice the demodulator cut-off frequency, plus some oversampling to account for finite filter attenuation in the stopband.
After downsampling, the time series blocks are concatenated and written to a compressed binary file in the HDF5 format \cite{HDF5} for later analysis.
Thanks to the streaming software architecture, the output signal can also be monitored graphically in quasi-real-time.
For this purpose, the time series blocks are buffered and re-sized for plotting in a graphical interface based on the Matplotlib library \cite{Matplotlib}.
Also, the noise power spectral density is computed by a standard Fourier periodogram using the \texttt{scipy.signal.welch} method and displayed in real time, giving immediate feedback on the noise properties of the DUT.

None of the signal processing steps produces any significant processor or memory load on a modern computer workstation, because the input sampling rates of order \(O(\SI{10}{\kilo\hertz})\) are far lower than the microprocessor clock frequency, and due to the efficiency of the filter algorithms.
Computationally expensive fast Fourier transforms (FFT) are only performed on the output signal, whose data rate has been further reduced by down-sampling.

Specific details about the software environment, installation and usage, as well as the implementation of new acquisition hardware are given in the \texttt{README.md} file of the software package, which is available for download at Ref.\ \cite{Thyzel2024Software}.
The modular code can be integrated into existing signal processing algorithms, see the ``Development'' section of that file. 

\section{Performance tests}

\begin{figure*}
\centering
\includegraphics[scale=1.0]{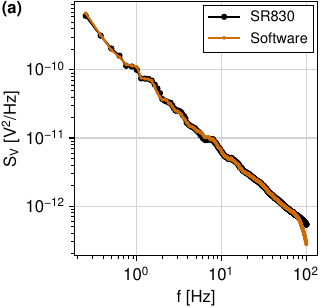}
\hspace{1mm}
\includegraphics[scale=1.0]{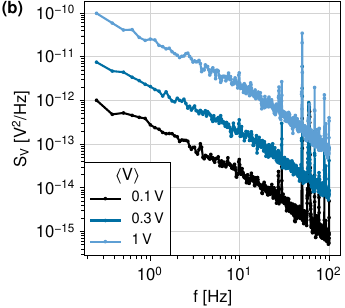}
\hspace{1mm}
\includegraphics[scale=1.0]{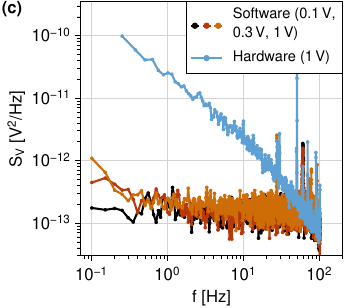}
\caption{
    \textbf{(a)}: Comparison of the voltage noise power density spectra of a real DUT exhibiting \(1/f\) noise, measured simultaneously with both methods.
    The spectra have been smoothed with a linear Savitzky-Golay filter (window width one tenth of a frequency decade) to better discern the quality of their match.
    \textbf{(b)}: Background voltage noise power density spectrum of the SR830 lock-in hardware for different RMS amplitudes \(\langle V \rangle\) of the pure sine wave input signal.
    \textbf{(c)}: Comparison of the background noise level of the software-based method for different values of \(\langle V \rangle\) to that of the SR830 hardware at \(\langle V \rangle = \SI{1}{\volt}\).}
\label{fig:noise-scaling}
\end{figure*}

First, a standard noise measurement with a real DUT was conducted using both the SR830 hardware, and the software lock-in amplifier with NI 9239, in order to verify the correct function and frequency response of the software implementation.
As the DUT, a bulk crystal of the organic charge-transfer salt \(\theta\)-(BEDT-TTF)\textsubscript{2}CsCo(SCN)\textsubscript{4} was chosen, which exhibits strong \(1/f\)-type resistance noise\cite{Thomas2022} that is further enhanced by introducing defects through X-ray irradiation \cite{Sasaki2012}.
The DUT resistance of \(\SI{3.6}{\kilo\ohm}\) was excited with a sine current \(I_\mathrm{rms} = \SI{20}{\micro\ampere}\) at a frequency of \(\SI{527}{\hertz}\), and the voltage drop was amplified with SR560 before being demodulated by SR830 and the software lock-in simultaneously.
The SR830 output noise spectrum was recorded using an SRS SR785 dynamic signal analyzer to verify the correctness of the spectrum calculation used in the software method.

Figure \ref{fig:noise-scaling}a shows that the power density spectra obtained by both methods match almost perfectly.
Indeed, the relative difference is less than about \(\SI{5}{\percent}\) over the whole frequency range, except at high frequencies, where the demodulator's low-pass filter depresses the spectrum of the software method.
Note that SR830 normally also exhibits this downturn, and in an even broader frequency range, because it employs cascaded R-C-type filters, which start to roll off far below the cutoff frequency (here \(\SI{160}{\hertz}\)).
We have corrected the SR830 spectrum for this filter response, but not that of the software solution. 

Having verified the frequency response, we next measured the background noise of both the SR830 hardware and the software lock-in implementation.
For this purpose, a noise-free device under test (DUT) was simulated by connecting the function generator DS360 directly to the inputs of SR830 and NI 9239 in parallel.
The lock-in amplifier instrument and the software-based solution therefore demodulated the same, un-amplified sinusoidal voltage signal with identical source and load impedances, and the same source noise level.
The circuit remained unchanged in all measurements with both devices.

Fig.\ \ref{fig:noise-scaling}b shows that SR830 exhibits intrinsic \(1/f\)-type noise with \(S_V(f = \SI{1}{\hertz}) \approx \SI{2e-11}{\per\hertz} \times \langle V \rangle^2\), whose magnitude depends quadratically on the input amplitude \(\langle V \rangle\).
This makes it difficult to distinguish from \(1/f\) noise originating from a DUT, which has the same scaling behaviour.
It also makes it impossible, without resorting to cross-correlation or bridge circuits suppressing \(\langle V \rangle\) \cite{Thyzel2024Methods}, to measure any DUT resistance noise smaller than \(S_R(\SI{1}{\hertz}) / R^2 = \SI{2e-11}{\per\hertz}\).
In contrast, the input noise of the software-defined lock-in setup shown in Fig.\ \ref{fig:noise-scaling}c does not exhibit any such dependence on the input amplitude.
At an input amplitude of \(\SI{1}{\volt}\), a significant reduction in background noise power density of about two orders of magnitude at \(f = \SI{1}{\hertz}\) is achieved.
In fact, the behavoiur of the software implementation resembles that of an ideal lock-in amplifier more closely, in that the input amplifier's \(1/f\) noise is frequency-shifted and therefore not visible.
This is not the case for the SR830 instrument, whose output exhibits very strong \(1/f\) noise.
As the signal at the inputs of both instruments was identical, SR830's undesirable \(1/f\) noise, which we verified under various other measurement conditions, must be caused either by the digital algorithms in its demodulator or the semiconductor components in its amplifier stages and ADC.
Note, however, that it is atypical for the input-referred noise level of simple amplifier or ADC circuits to increase with the input amplitude like in Fig.\ \ref{fig:noise-scaling}b.
In conclusion, it remains unknown what causes the inconvenient \(1/f\) noise in SR830, but our implementation of the lock-in method can offer superior low-frequency noise performance, in particular when multi-stage amplifiers are used to exploit the higher input range of \(\pm\SI{10}{\volt}\) for NI 9239.

\begin{figure}
\centering
\includegraphics[scale=1.0]{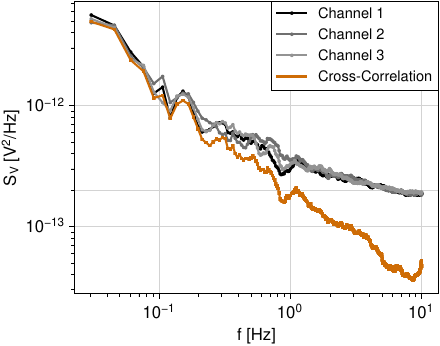}
\caption{Lower-frequency measurement of the noise background of the software lock-in method for \(\langle V \rangle_\mathrm{rms} = \SI{1}{\volt}\), analogous to Fig.\ \ref{fig:noise-scaling}c.
    The orange curve shows the mean of the cross power spectral densities of all channel combinations, i.e.\ \(\lvert S_{12}(f) + S_{13}(f) + S_{23}(f) \rvert / 3\).
    Spectra have been smoothed as in Fig.\ \ref{fig:noise-scaling}a.
}
\label{fig:background}
\end{figure}
Finally, we investigated the noise background of the software lock-in implementation at very low frequencies in more detail.
For this purpose, we again fed a sine wave directly into the NI 9239 channels.
Three input channels were demodulated independently, but sharing the same reference channel.
The noise spectrum shown in Figure \ref{fig:background} exhibits an \(1/f\)-like upturn at extremely low frequencies corresponding to a time scale of seconds.
Surprisingly, this part of the spectrum seems to be fully correlated between the three input channels, as the cross power spectral density \cite{Sampietro1999} computed by \texttt{scipy.signal.csd} matches the single-channel spectrum.
This indicates that the \(1/f\) contribution is caused either by the common reference channel, or by the sine signal source.
We speculate that in the latter case, noise in the oscillator frequency driving the signal generator output could cause a \(1/f\)-like modulation of the sine signal.
In any case, the \(1/f\) noise magnitude is only approximately \(S_V(f = \SI{1}{\hertz})/\langle V \rangle^2 \approx \SI{2e-13}{\per\hertz}\), two orders of magnitude lower than the \(1/f\) background of the SR830 instrument.

\section{Conclusion and outlook}

We have implemented a software-defined lock-in amplifier for use in amplitude-modulated low-frequency resistance noise measurements.
Compared to previous implementations, our streaming software architecture allows for highly efficient signal processing with real-time preview capabilities, utilizing commercially available data acquisition hardware.
We show the correctness of the software implementation by comparing its output noise spectrum with that of the commercial, high-end lock-in amplifier instrument Stanford Research Systems SR830.
Importantly, our implementation exhibits a \(1/f\) noise power spectral density at \(\SI{1}{\hertz}\) about two orders of magnitude below that of the commercial instrument.
This makes it possible to investigate devices with an extremely low noise level, which could previously only be detected in bridge circuit configurations that are not available for all device types.
Future improvements could include scaling up the number of DAQ channels and pre-amplifiers in a cross-correlation setup to suppress the remaining noise background further, as suggested in Ref.\ \cite{Scandurra2013}.
Furthermore, multiple reference channels, instead of a single one, could possibly be used to reduce the residual \(1/f\) contribution.
Aside from applications in noise spectroscopy, a dual-phase detection using a software-based phase-locked loop as in Ref.\ \cite{Uhl2021} would allow our solution to be used for complex impedance spectroscopy at low frequencies.

\section{Acknowledgements}

I thank Professor Jens Müller (Physikalisches Institut, Goethe-Universität, Frankfurt am Main, Germany) for providing the resources for and supporting this project, as well as valuable discussions.
Funding was provided by the Deutsche Forschungsgemeinschaft (DFG) through SFB/TRR288 (422213477), project B02.

\bibliography{paper}
\end{document}